\documentclass{CIRED-v1}
\usepackage{soul}
\usepackage{comment}
\usepackage[hidelinks]{hyperref}
\usepackage{xcolor}

\begin{document}

\title{WEcharge: democratizing EV charging infrastructure}

\author{Md Umar Hashmi\ad{1\corr}, Mohammad Meraj Alam\ad{2}, Ony Lalaina Valerie Ramarozatovo\ad{3}, Mohammad Shadab Alam\ad{4}\ 
}

\address{\add{1}{KU Leuven \& EnergyVille, Belgium}
\add{2}{FlexThor BV, Belgium}
\add{3}{Independent Consultant, France}
\add{4}{University of Hasselt, Belgium}
\email{mdumar.hashmi@kuleuven.be}}


\begin{abstract}
The sustainable growth of EVs will have to be met with proportional growth in EV charging infrastructure. With limited urban spaces to place new charging stations, shrinking profitability, privately owned charging facilities need to be shared. WEcharge will allow privately owned charging infrastructure to be shared with public EV owners using a business model.
We propose a resource matching algorithm that takes into account incoming EV preferences, hard constraints for such EV, and provides the best suited resource for charging.
We demonstrate the applicability of the matching model by showing a realistic case study with a Nissan Leaf 40 kW EV and 25 company and publicly owned charging stations (DC fast charger, AC rapid charger, level 1 and level 2 charger) in Hasselt, Belgium. The case study shows that consumer preferences will govern resource matching.
\end{abstract}

\maketitle

\section{Introduction}
\begin{figure}[!htb]
\centering\includegraphics[width=0.96\linewidth]{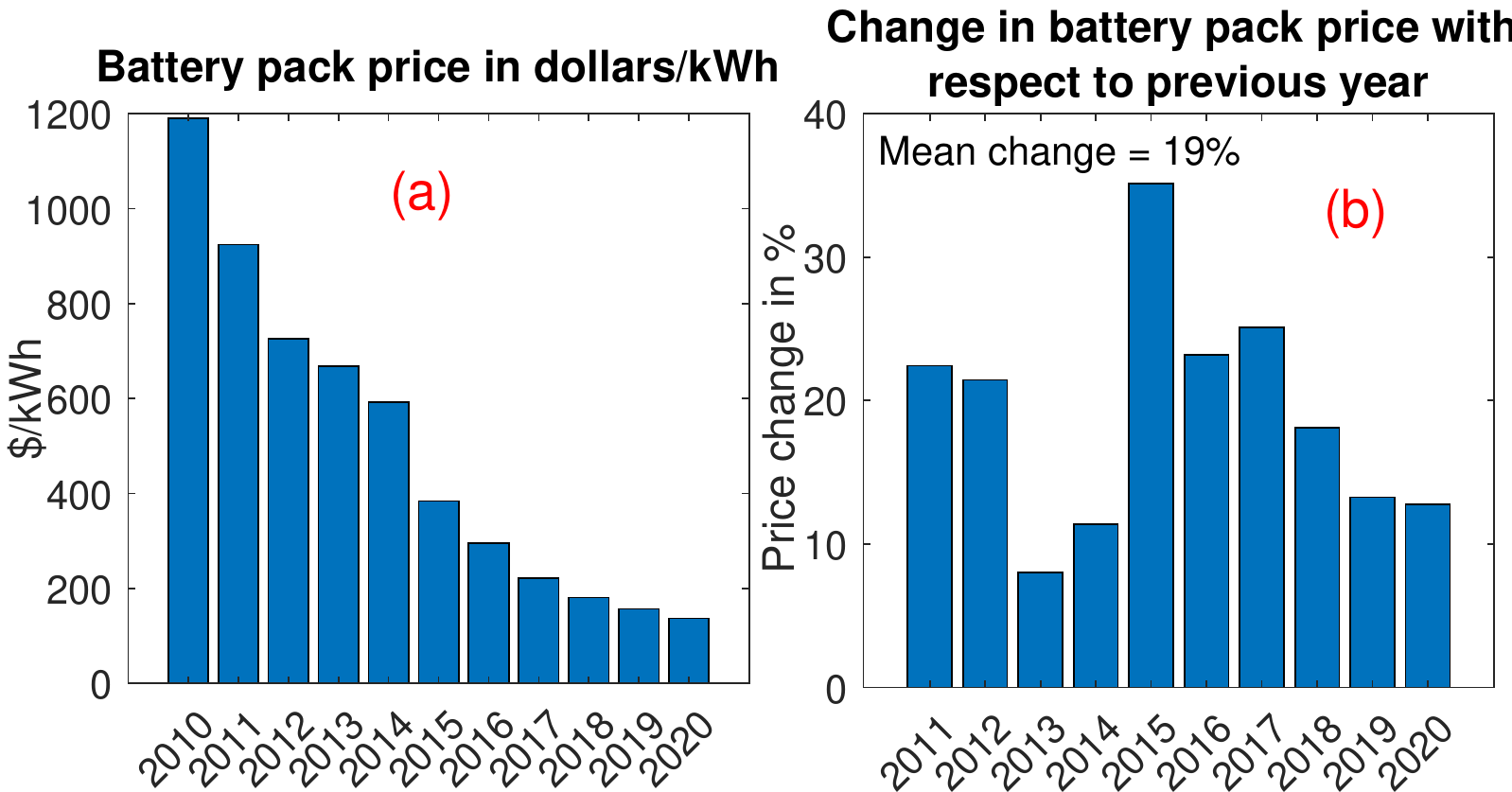}\vspace{-7pt}
\caption{(a) Lithium battery pack price (b) Mean change in battery pack prices \cite{bloomberg}} 
\label{fig1}\vspace*{-12pt}
\end{figure}

Electric vehicle (EV) share in transportation is expected to grow in the coming decades.
According to BNEF report, every sixth car sold in the world will be an EV by 2025 and the global sales can hit approximately 17 million \cite{bloomberg}. Although, a slow growth in the first quarter of 2020 and 2019 has been observed, even then the overall market of EV sales rose to 26 percent and overall by 44 percent in Europe in 2021 \cite{mckinsey}. Moreover, US Energy Information Administration (EIA) predicts that by 2050, EV fleet will compose 34\% in OECD countries and 28\% of total light duty vehicles in non-OECD countries \cite{euobservatory}. This projection is from EV consisting of a meagre 0.7\% of the total light duty vehicles in the year 2020 to 31\% globally by 2050 \cite{eiaprojects}. One of the main reasons behind this rapid growth and market acceptability of EVs has been the steep decline in lithium battery prices over the last decade, with a mean change of 19\% year-on-year, as shown in Figure \ref{fig1} (a) and (b).

\begin{figure}[!htb]
\centering\includegraphics[width=0.8\linewidth]{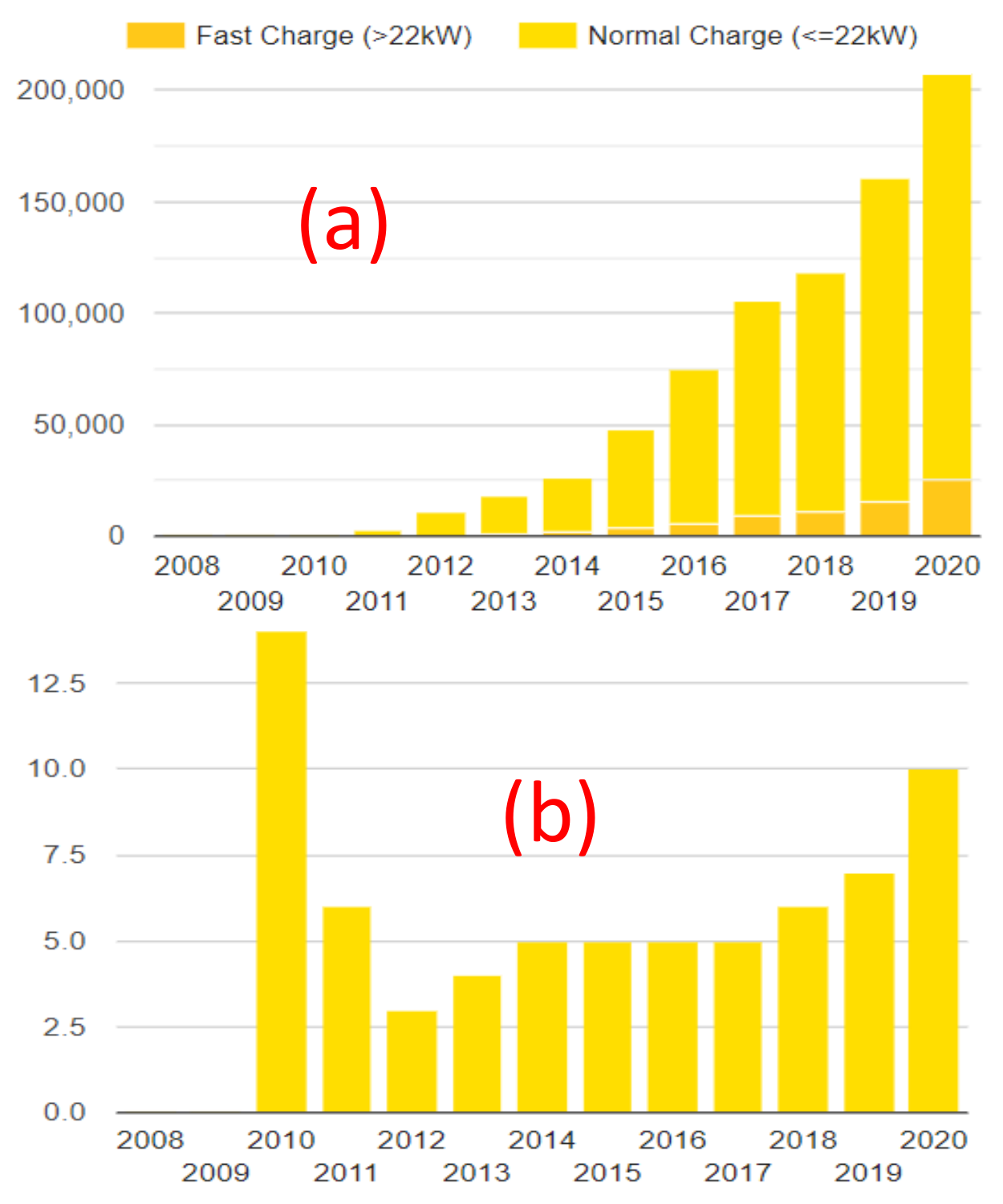} \vspace{-7pt}
\caption{(a) Growth of EV charging infrastructure in Europe (b) Average EV per public charging facility \cite{euobservatory}.}
\label{fig2}\vspace*{-12pt}
\end{figure}

The unprecedented growth of EV is helping the market and the governments to achieve the zero emissions objective. However. with such an unparalleled growth of EVs, the stress on existing EV charging infrastructure is growing as well, leading to an increase in government and company owned investments in installing new charging infrastructure \cite{hall2017emerging}. 
Figure \ref{fig2} (a) and (b) shows the growth of EV charging infrastructure in Europe and on average EV per public charging facility. {As EVs become more popular, more public charging stations need to be installed which is expensive, time-consuming, permit required and grid connections upgrades. According to a study done by NREL, the US will need to build 50000 DC fast chargers and 1.2 million level 2 fast chargers for 35 million EVs by 2030 \cite{wood2017national}.}

Many EV owners have installed EV chargers at their homes which they use to charge their EV’s generally after the office hours (in the evening and at night when the network tariff is low) and during the weekends. Since these privately-owned chargers are available only for private use, they remain non-utilized from 16 to 18 hours per day on an average (60\% – 70\% of duration) \cite{virta}. The reason behind this non-participation of privately-owned chargers with normal public is non-existence of any digital platform at this moment where the homeowners can host their privately-owned chargers on rent and monetize their energy asset when not in use. There is a clear gap between the private charger owner and public EV owner, where a business landscape exist to connect both players together.

This paper discusses the pain point of EV owners when travelling to a destination with incomplete real-time information about availability of public charging stations during peak-hours, under-utilized privately-owned chargers, and presents an innovative business model between the privately-owned charger owner and EV owner to fill this gap in section \ref{section2}. In section \ref{section3}, an adaptive infrastructure matching algorithm is proposed to match charging stations with incoming EVs. Moreover, a case study on matching mechanism has been demonstrated with a 40 kW Nissan Leaf and 25 charger stations in a city of Belgium, Hasselt, in section \ref{section4}. Section \ref{section5} discusses the conclusions of the paper.


\section{Business model}
\label{section2}
Presently, an EV owner has incomplete real-time information about any EV charging station infrastructure such as (i) if a resource occupied or idle, (ii) charger and socket type, (iii) location, (iv) charging tariffs and penalties, etc.
When EV owners travel to other destination, they charge their EV’s using charging stations owned by the government or charging stations companies. According to a survey, there is a waiting time of 10 minutes to 15 minutes for EV owners in real-time to find a publicly available charging spots during the peak-hours while many private charging stations owned by homeowners, SME’s and private companies stand idle and underutilized \cite{evcharg}.

The goal of WEcharge platform is to
reduce and evenly distribute (increasing utilization factor of underutilized resources) the stress on the EV charging station infrastructure by bringing the homeowner and private companies owned charging station into the mainstream through a privately-owned charging station-as-a-service business model. Using our platform, not only the public charging stations owned by the government will be listed but also the charging stations owned by any private companies and homeowners can be listed along with their characteristics as
\begin{itemize}
    \item 	real-time and future availability,
    \item	charging tariff and mode of charging,
    \item	charger type, socket, number of phases, and AC/DC,
    \item	electrical properties, voltage, power rating, amperage,
    \item	geographical coordinates,
    \item	manufacturer and mode of payment.
\end{itemize}
Private charging stations owned by homeowners, SME’s and private companies can participate in this business model to decrease their charger non-utilization time and earn remuneration to decrease the payback of their installation. Moreover, a service to reserve a charging spot in advance or pre-book any charging stations before commencing any travel is facilitated. Additionally, the platform notifies EV owner in case of penalty for not removing EV after its fully charged.
\subsection{Customer segment}
The WEcharge rent-a-charging spot service directly focuses on the following customer segments, where the EV owner receives the ease of charging facility at the tip of his finger and the other hand charger owner earns money on their under-utilized privately-owned chargers. The EV charging infrastructure includes
\begin{itemize}
    \item 	charging stations privately owned by privately,
    \item	charging stations owned by SME’s and companies,
    \item	and public/municipal owned charging stations.
\end{itemize}
\subsection{Revenue model and partnerships}
The platform services are a subscription based model that includes services like pre-book charging stations, EV battery monitoring, real-time pricing, price forecasting and charger-as-a-service participation. The customers have the freedom to choose pay-as-you-go model with higher subscription fee as the services increases. There are potential partnerships as shown in Fig. \ref{fig4} listed as
\begin{itemize}
    \item 	DSO and TSO for improving grid stability,
    \item	EV charger OEMs for up-to-date standards and protocols,
    \item	Charger installers for upgrading charging infrastructure,
    \item	Municipal/government for policies and regulation updates.
\end{itemize}
\vspace{-8pt}
\begin{figure}[!htb]
\centering\includegraphics[width=0.96\linewidth]{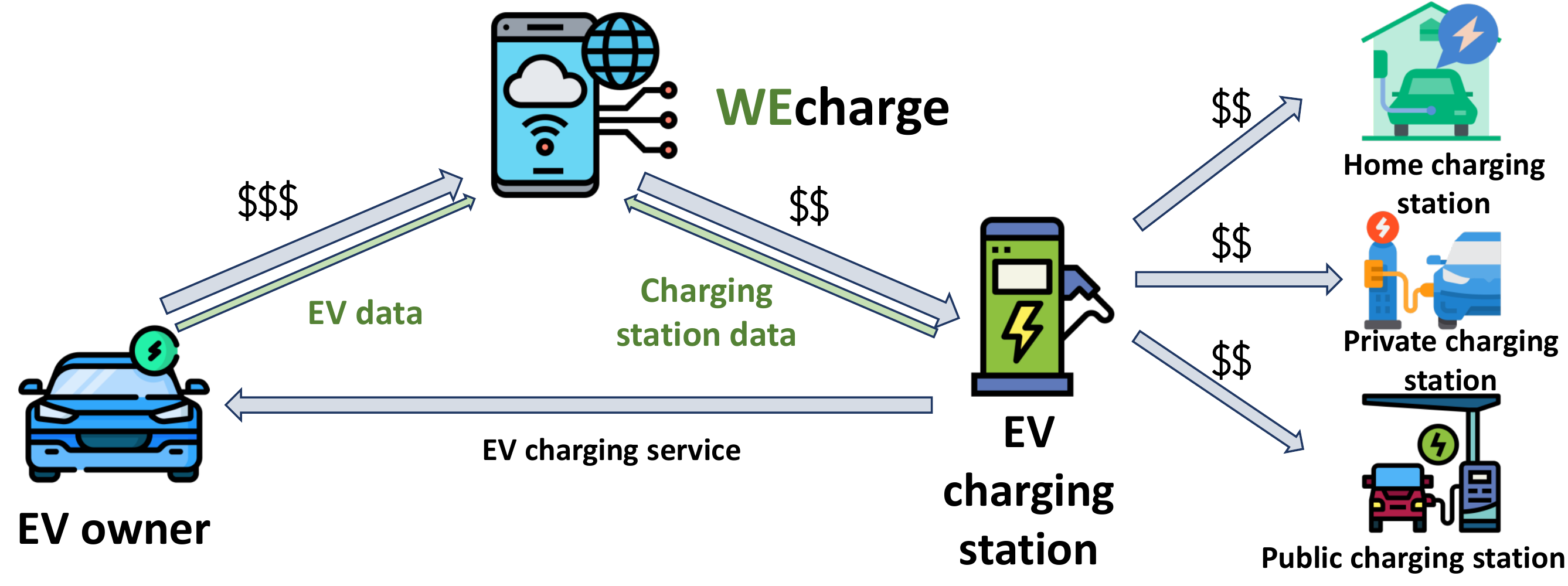}  \vspace{-7pt}
\caption{WEcharge business model}
\label{fig3}\vspace*{-12pt}
\end{figure}

\begin{figure}[!htb]
\centering\includegraphics[width=0.95\linewidth]{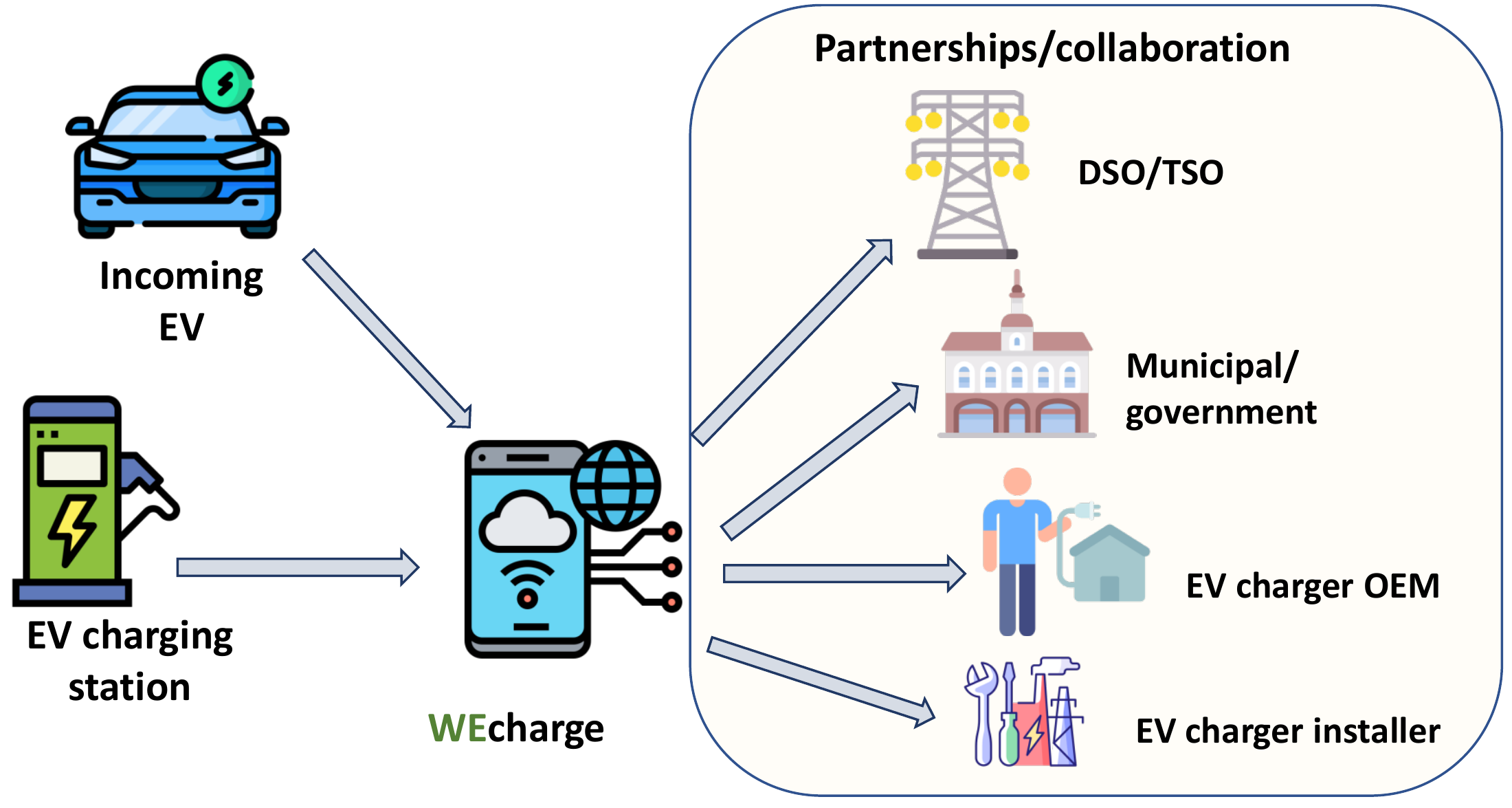}  \vspace{-8pt}
\caption{WEcharge business partnerships}
\label{fig4}\vspace*{-12pt}
\end{figure}

\section{WEcharge: adaptive resource matching}
\label{section3}
{
The increase in EVs will make it more challenging to install new charging infrastructure due to space constraint, and shrinking profitability \cite{flowbank}. This may bottleneck future growth of EVs \cite{forbes9}.}
The increased investment could be avoided by facilitating private EV charging infrastructure to share their resources for charging incoming EVs. This will reduce stress on public and company owned charging resources, thus avoiding a probable bottleneck in the growth of EV. Similar to Uber, which matches taxi driver to interested rider, our proposed solution, WEcharge, matches incoming EVs to a pool of public and private charging infrastructures based on (a) charging preference, (b) EV make, (c) geographical proximity, (d) travel time to charging facility, (e) remaining state-of-charge (SoC) of EV battery, (f) cost of charging, (g) reduce incoming EV waiting period. Additional criteria for matching can be added for considering consumer preferences, local distribution network congestion issues, renewable energy integration, driving range, health of EV batteries \cite{hall2017emerging} etc.

\begin{figure}[!htb]
\centering\includegraphics[width=0.96\linewidth]{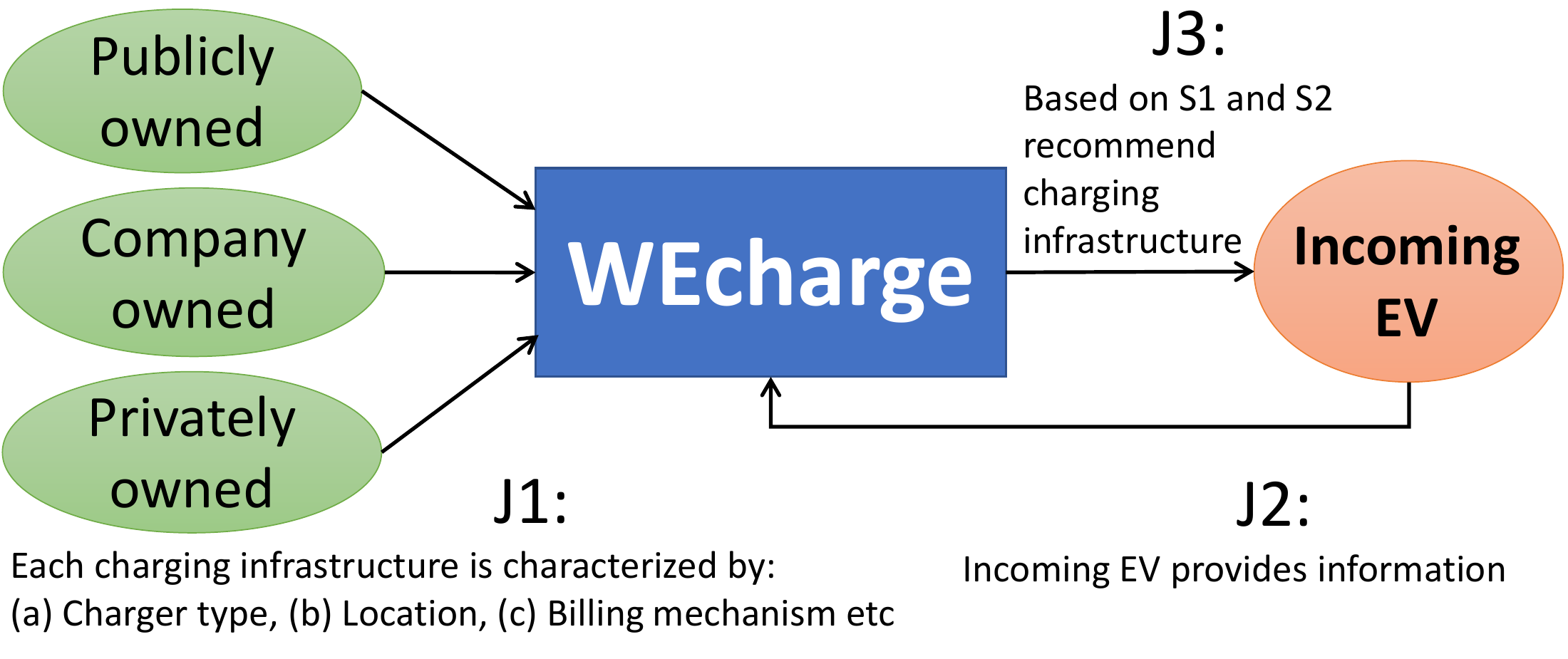}  \vspace{-7pt}
\caption{Matching incoming EV to charging infrastructure.}
\label{fig5}\vspace*{-12pt}
\end{figure}

\begin{figure}[!htb]
\centering\includegraphics[width=0.45\linewidth]{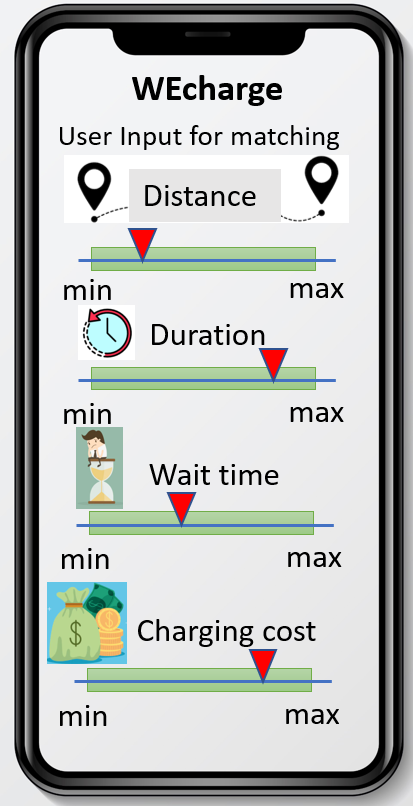}  \vspace{-7pt}
\caption{The incoming EV charging preference is shown. These weights are used for solving \eqref{matcheq}.}
\label{fig6}\vspace*{-12pt}
\end{figure}

WEcharge will work autonomously and does not require human intervention. However, in case the private charging infrastructure is going to be not available, the owner can manually opt out. This will result in dropping of this resource from WEcharge platform. This resource must be locally activated to be shown in WEcharge.
The skeleton of the matching model is shown in Fig. \ref{fig5}.
{
In order to utilize the WEcharge platform, in step (J1) it gathers attributes of the EV chargers, in step (J2) the incoming EV provides its charging preferences as shown in Fig. \ref{fig6}, and the matching algorithm is executed in (J3) step to find the best suited EV charger meeting consumer needs. }
The bottleneck for WEcharge is the lack of universal EV charging infrastructure: this could stress resources more than others, depending on the growth of certain types of EVs. For example, if Tesla EVs compose of 30\% total EVs, however, Tesla compatible chargers compose of 40\% then on average all Tesla compatible charging infrastructure will be underutilized. 

In Belgium, new charging infrastructure placement requires a feasibility study conducted by an electric utility for analysing the impact of charging stations on the distribution network. It is probable that some location with higher probabilities of voltage and thermal congestion may limit such a placement. However, such nodes will be highly effective in mitigating probable congestion issues on such and neighbourhood nodes. Incoming EVs which are flexible in charging profile could significantly contribute in realization of active distribution network.

Proposed matching models have the following properties:
\begin{itemize}
\item the democratization of charging infrastructure so as anyone can participate. WEcharge encourages privately-owned EV charging infrastructure to participate in resource sharing thus
	avoids public and corporate investment in new charging infrastructures, and
    \item	matches incoming EV with charging infrastructure while considering various charging modes and preferences.
\end{itemize}

\subsection{Resource matching algorithm}
The resource matching algorithm classifies the vehicle parameters and user inputs into hard and soft constraints.
The hard limitations need to be satisfied and soft limitations should be optimized to find the best alternative.

The \textit{hard constraints} include  
\begin{itemize}
    \item Vehicle charger type and socket compatibility,
    \item Will the remaining SOC in the electric vehicle be able to reach the charging facility or not, and
    \item Availability of charging infrastructure at the present time or for pre-booking.
\end{itemize}

The \textit{soft constraints} include
\begin{itemize}
    \item Geographical distance denotes the distance between vehicle and the charging infrastructure,
    \item Duration of charging at the rated charging rate,
    \item Waiting time: is assumed to be inversely proportional to the number of chargers at the charging station,
    \item Cost of charging: assuming EV is charged from zero SoC.
\end{itemize}

Consider there are $N$ charging infrastructures which satisfies the hard constraints. An option $O_i$ for resource $i$ is given as
\begin{equation}
    O_i = \{ d_i, t_i, s_i, c_i\},
\end{equation}
where $d_i$ is the geographical distance, $t_i$ is the charging duration, $s_i$ denotes waiting time, and $c_i$ is the cost of charging.
{Note, in this paper we present a simple matching model. Several additional considerations can also be included.}

The performance metric is denoted as 
\begin{equation} 
\begin{split}
P_i  & = \Big( \frac{w_1}{\bar{w}} \frac{d_i}{\max_{i\in \{1,..,N\}} d_i} +
    \frac{w_2}{\bar{w}} \frac{t_i}{\max_{i\in \{1,..,N\}} t_i}  \\
 & +\frac{w_3}{\bar{w}} \frac{s_i}{\max_{i\in \{1,..,N\}} s_i} +
    \frac{w_4}{\bar{w}} \frac{c_i}{\max_{i\in \{1,..,N\}} c_i} \Big),
\end{split}
\end{equation}


The matching algorithm returns the best suited charging infrastructure, given as
\begin{equation} \label{matcheq}
i^*   = \arg \min_{i} P_i,
\end{equation}
where $w_1, w_2,w_3, w_4$ denotes the weights for features $d_i, t_i, s_i, c_i$ respectively. These weights are input by a user looking for matching infrastructure. 
$\bar{w}$ denotes the mean weights, given as
$
    \bar{w} = \frac{1}{4} \sum_{k=\{1,2,3,4\}} w_k.
$

In \eqref{matcheq}, the weights and features are normalized for avoiding the skewness caused by magnitude difference of the feature parameters.
For all the parameters in $O_i$ needs to minimized, therefore, \eqref{matcheq} is a sum of product of normalized weights and normalized features to be minimized. In case of a parameter which needs to be maximized, such as utility of EV owner, then it should be subtracted in \eqref{matcheq}.

Note that one charging infrastructure can be considered more than an option based on number of modes of charging. Some of the EV charging modes could be
\begin{itemize}
    \item Only charging,
    \item Vehicle to grid mode of charging:
    \begin{itemize}
        \item Providing services to DSO by performing voltage regulation, congestion management, peak reduction etc.,
        \item Providing services to TSO by performing ancillary services, frequency regulation etc.,
        \item Performing energy arbitrage or power quality improvement based on billing rules of the retailer.
    \end{itemize}
\end{itemize}
The details of these modes of charging are not included in this paper. 
Since the proposed matching algorithm is computationally fast, therefore, real-time adaptive mapping will be easily implemented. This feature will be desired by moving EVs searching for an electric vehicle charging point.

In the next section, we detail a case study showing how this algorithm can be applied for real world mapping.

\section{Case-study}
\label{section4}
The case study demonstrates the matching mechanism proposed in this paper.
Table \ref{tab:incoming} lists the details of the incoming EV.
We consider EV charging matching of a Nissan Leaf (2018) 40 kW located at point $x$, shown in Fig. \ref{fig6}.
Based on the remaining SoC of the battery, suppose 25 charging stations are present, as shown in Fig. \ref{fig6}.
Note that except for charger 15 in Fig. \ref{fig6}, all other chargers are compatible to Nissan Leaf, implying hard constraints are satisfied.

\begin{table}
\centering
\begin{center}
\caption {Incoming EV details \cite{podpoint, mobilityhouse, evbox}}
\label{tab:incoming}
\begin{tabular}{|l|l|l|} 
\hline
\multicolumn{2}{|l|}{Model}                    & Nissan Leaf 2018  \\ 
\hline
\multicolumn{2}{|l|}{Battery capacity}         & 40kWh             \\ 
\hline
\multicolumn{2}{|l|}{Total electrical range}   & 378 km            \\ 
\hline
\multicolumn{2}{|l|}{Plug Type}                & Type 2            \\ 
\hline
Domestic, 10 A, 1-phase & 2.3 kW               & 17 hours          \\ 
\hline
1-Phase, 16A            & 3.7 kW               & 11 hours          \\ 
\hline
1-Phase, 32A            & 7.4 kW               & 6 hours           \\ 
\hline
3-Phase, 16A            & 11 kW                & 10 hours          \\ 
\hline
3-Phase, 32A            & 22 kW                & 6 hours           \\ 
\hline
DC Fast Charger         & 50 kW                & 45 minutes        \\
\hline
\end{tabular}
\end{center}
\end{table}

\begin{figure}[!htb]
\centering\includegraphics[width=1\linewidth]{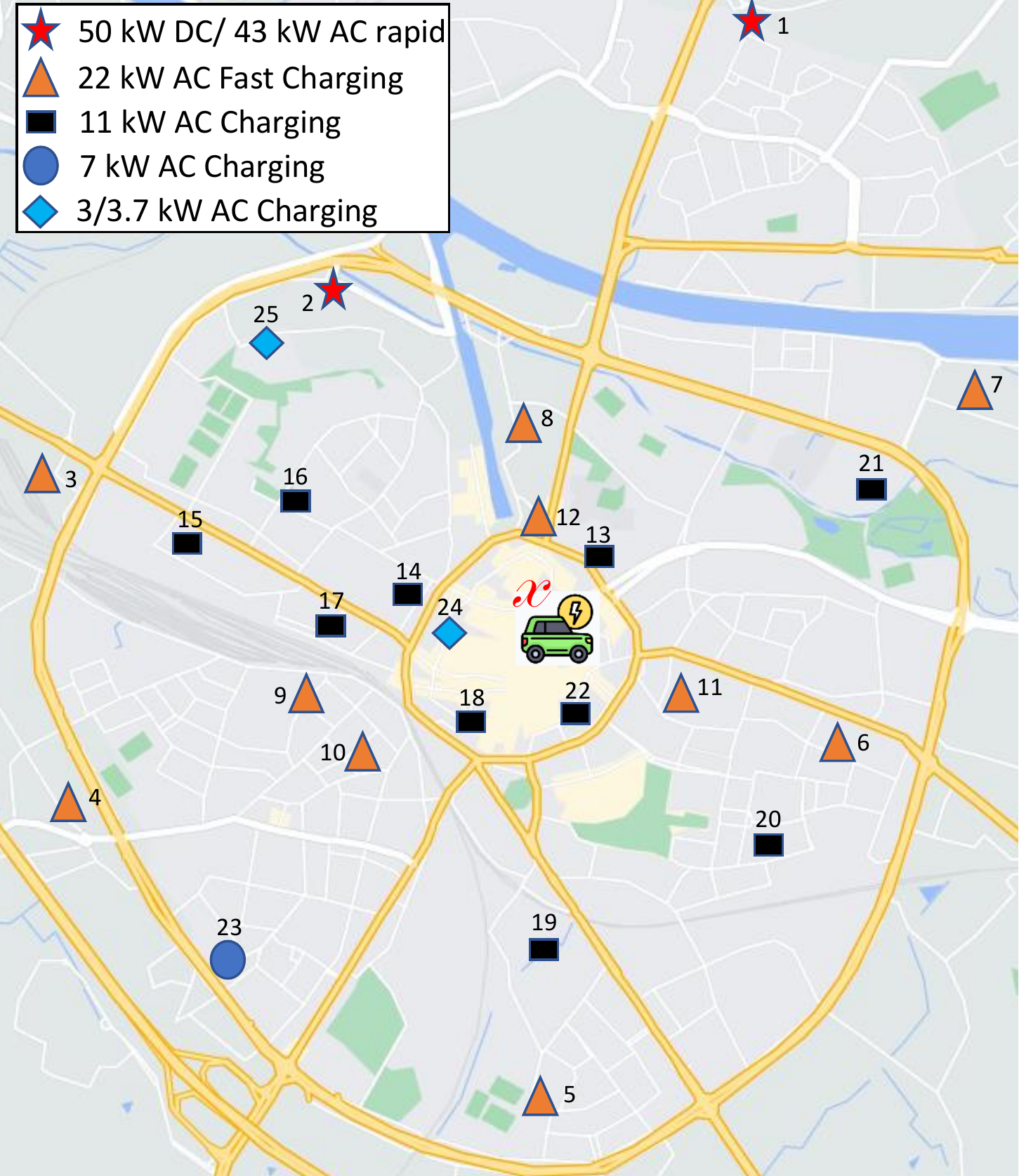}
\caption{Map of Hasselt in Belgium \cite{google, chargemap}, with EV charging infrastructure marked (non-exhaustive list). The location of incoming EV is marked with $x$.}
\label{fig7}\vspace*{-12pt}
\end{figure}

\begin{figure}[!htb]
\centering\includegraphics[width=0.9\linewidth]{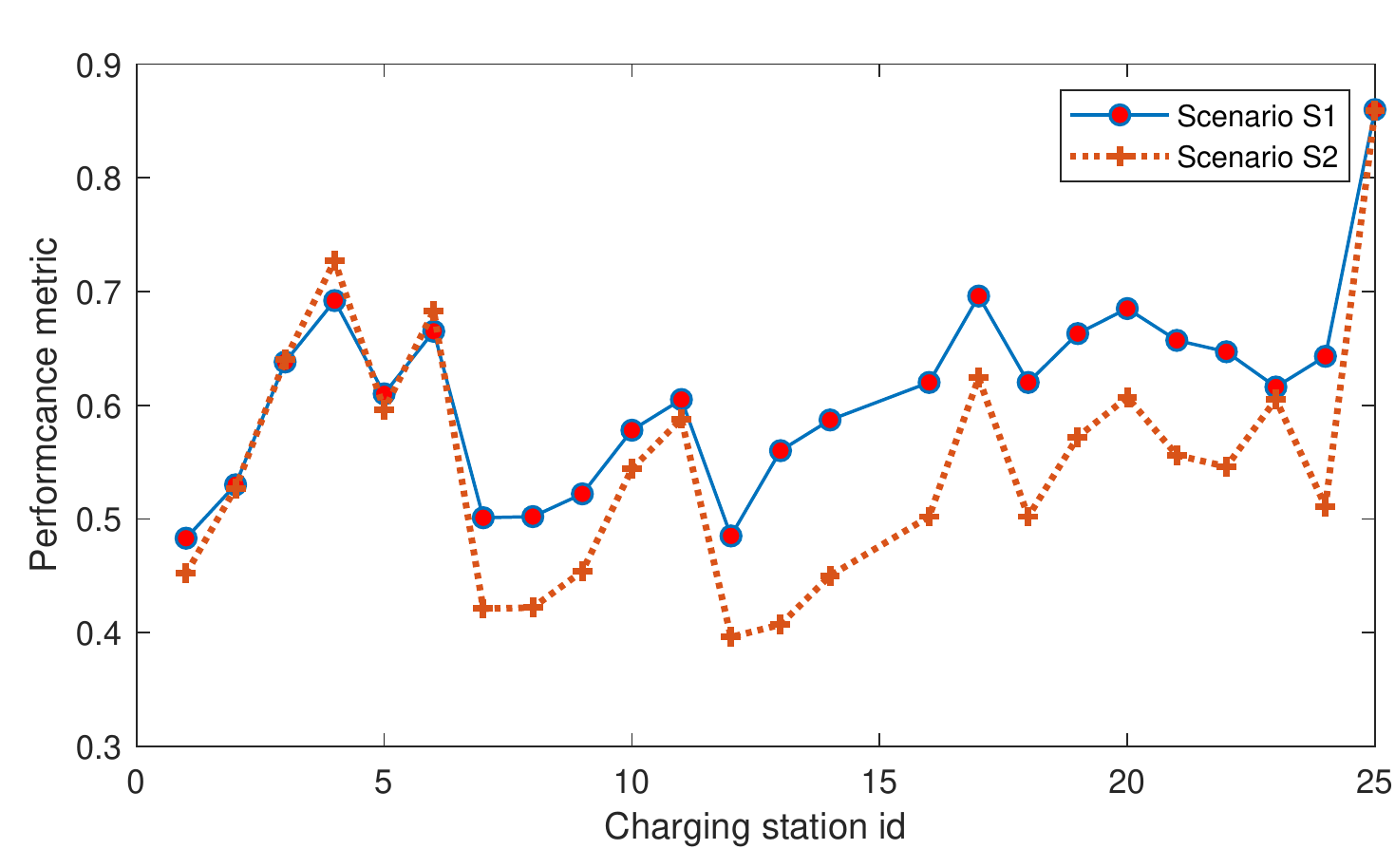}  \vspace{-10pt}
\caption{Performance metrics}
\label{fig8}\vspace*{-12pt}
\end{figure}

Table \ref{tab:perfindex} summarizes the performance metrics based on four parameters described in \eqref{matcheq}.
The performance metrics are calculated for two scenarios:
\begin{itemize}
    \item S1: Equal weights; $w_1=w_2=w_3=w_4=0.25$,
    \item S2: Unequal weights; $w_1=w_3=0.4$, and $w_2=w_4=0.1$.
\end{itemize}
Fig. \ref{fig8} shows the performance metric for all charging infrastructure satisfying the hard constraints.
For scenario S1 the first EV station is best suited and for S2 EV charging station id 12 should be selected.
Clearly, we can observe that incoming EV charging preferences will govern the selection of charging infrastructure.
\begin{table}
\centering
\caption {Performance metrics for vehicle at point $x$}
\label{tab:perfindex}
\begin{tabular}{|l|l|l|l|l|l|l|}
\hline
  &  &  &     &    Norm.     & S1 & S2    \\
id & Wait  & Norm.    & Norm. & charge      & Equal & Varying  \\
   & time  & distance & cost  & time & weight  & weight   \\
   \hline
1  & 0.17 & 0.696    & 1     & 0.068       & \hl{0.483} & 0.452    \\
2  & 0.33 & 0.717    & 1     & 0.068       & 0.530 & 0.527    \\
3  & 0.5   & 0.783    & 0.722 & 0.545       & 0.638 & 0.640    \\
4  & 0.5   & 1        & 0.722 & 0.545       & 0.692 & 0.727    \\
5  & 0.5   & 0.674    & 0.722 & 0.545       & 0.610 & 0.596    \\
6  & 0.5   & 0.891    & 0.722 & 0.545       & 0.665 & 0.683    \\
7  & 0.08 & 0.652    & 0.722 & 0.545       & 0.501 & 0.421    \\
8  & 0.5   & 0.239    & 0.722 & 0.545       & 0.502 & 0.422    \\
9  & 0.17 & 0.652    & 0.722 & 0.545       & 0.522 & 0.454    \\
10 & 0.5   & 0.543    & 0.722 & 0.545       & 0.578 & 0.544    \\
11 & 0.5   & 0.652    & 0.722 & 0.545       & 0.605 & 0.588    \\
12 & 0.5   & 0.174    & 0.722 & 0.545       & 0.485 & \hl{0.396}    \\
13 & 0.5   & 0.109    & 0.722 & 0.909       & 0.560 & 0.407    \\
14 & 0.5   & 0.217    & 0.722 & 0.909       & 0.587 & 0.450    \\
16 & 0.5   & 0.348    & 0.722 & 0.909       & 0.620 & 0.502    \\
17 & 0.5   & 0.652    & 0.722 & 0.909       & 0.696 & 0.624    \\
18 & 0.5   & 0.348    & 0.722 & 0.909       & 0.620 & 0.502    \\
19 & 0.5   & 0.522    & 0.722 & 0.909       & 0.663 & 0.572    \\
20 & 0.5   & 0.609    & 0.722 & 0.909       & 0.685 & 0.607    \\
21 & 0.5   & 0.478    & 0.741 & 0.909       & 0.657 & 0.556    \\
22 & 0.5   & 0.457    & 0.722 & 0.909       & 0.647 & 0.546    \\
23 & 0.5   & 0.696    & 0.722 & 0.545       & 0.616 & 0.605    \\
24 & 0.5   & 0.348    & 0.722 & 1           & 0.643 & 0.511    \\
25 & 1     & 0.717    & 0.722 & 1           & 0.860 & 0.859 \\
\hline
\end{tabular}
\end{table}

\section{Conclusion}
\label{section5}
Democratization of EV infrastructure is a crucial next step to ensure sustained growth of new EVs.
We propose a business model, WEcharge, which could allow privately owned EV charging infrastructure to share their resource. This will reduce congestion and waiting time for company and publicly owned charging infrastructure.
WEcharge will also assist in mapping incoming EVs to their best suited EV charging infrastructure. We demonstrate the applicability of this matching algorithm using a realistic case study. This case study consider Renault Leaf at the city centre in Hasselt in Belgium. There are 25 options for charging. Our matching algorithm provides the best suited charging infrastructure based on consumer preferences.

In future work, we will extend this matching models to consider more real-time parameters and various modes of charging while providing services to DSOs and/or TSO.

\section*{References}
\bibliographystyle{IEEEtran}
\bibliography{reference}

\end{document}